\documentclass[prl,aps,twocolumn,showpacs,preprintnumbers,amsmath,amssymb,floatfix]{revtex4}

\usepackage[dvips]{graphics,graphicx,color}

\begin{document}

\title{Quantum Monte Carlo Study of a Positron in an Electron Gas}

\author{N.\ D.\ Drummond}

\affiliation{Department of Physics, Lancaster University, Lancaster LA1 4YB,
United Kingdom}

\author{P.\ L\'opez\ R\'{\i}os}

\affiliation{TCM Group, Cavendish Laboratory, University of Cambridge, J.\ J.\
  Thomson Avenue, Cambridge CB3 0HE, United Kingdom}

\author{C.\ J.\ Pickard}

\affiliation{Department of Physics, University College London, Gower Street,
London WC1E 6BT, United Kingdom}

\author{R.\ J.\ Needs}

\affiliation{TCM Group, Cavendish Laboratory, University of Cambridge, J.\ J.\
  Thomson Avenue, Cambridge CB3 0HE, United Kingdom}

\date{\today}

\begin{abstract}
  Quantum Monte Carlo calculations of the relaxation energy, pair-correlation
  function, and annihilating-pair momentum density are presented for a
  positron immersed in a homogeneous electron gas.  We find smaller relaxation
  energies and contact pair-correlation functions in the important low-density
  regime than predicted by earlier studies.  Our annihilating-pair momentum
  densities have almost zero weight above the Fermi momentum due to the
  cancellation of electron-electron and electron-positron correlation effects.
\end{abstract}


\maketitle

Electron-positron annihilation underlies both medical imaging with positron
emission tomography (PET) and studies of materials using positron annihilation
spectroscopy (PAS) \cite{krause_rehberg}.  Positrons entering a material
rapidly thermalize and the majority annihilate with opposite-spin electrons to
yield pairs of photons at energies close to 0.511 MeV\@.  In a PET scan,
positrons are emitted by radionuclides in biologically active tracer molecules
and the resulting annihilation radiation is measured to image the tracer
concentration.  The interaction of low-energy positrons with molecules is
therefore of substantial experimental and theoretical interest
\cite{Gribakin_2010}.  PAS is used to investigate microstructures in metals,
alloys, semiconductors, insulators \cite{krause_rehberg}, polymers
\cite{Pethrick_1997}, and nanoporous materials \cite{Gidley_2006}.  Positrons
are repelled by the positively charged nuclei and tend to become trapped in
voids within the material.  The positron lifetime is measured as the interval
between the detection of a photon emitted in the $\beta^+$ radioactive decay
that produces the positron and the detection of the annihilation radiation
\cite{krause_rehberg}.  The lifetime is characteristic of the region in which
the positron settles, and PAS is a sensitive, nondestructive technique for
characterizing the size, location, and concentration of voids in materials.
Measuring the Doppler broadening of the annihilation radiation or the angular
correlation between the two 0.511 MeV photons yields information about the
momentum density (MD) of the electrons in the presence of the positron.  These
techniques may be used to investigate the Fermi surfaces of metals
\cite{Major_2004}.

The aim of PAS experiments is to investigate a host material without the
changes induced by the positron.  The positron is, however, an invasive probe
which polarizes the electronic states of the material.  Disentangling the
properties of the host from the changes induced by the positron is a major
theoretical challenge.  Positrons in condensed matter may be modeled with
two-component density functional theory (DFT) \cite{boronski_1986}, in which
the correlations are described by a functional of the electron and positron
density components.  Within the local density approximation (LDA), this
functional is obtained from the difference $\Delta \Omega$ between the energy
of a homogeneous electron gas (HEG) with and without an immersed positron.
$\Delta \Omega$ is known as the relaxation energy, and is equal to the
electron-positron correlation energy.

Two-component DFT gives reasonable electron and positron densities, but the
DFT orbitals do not describe electron-positron correlation properly
\cite{boronski_1986,puska_1994}.  The electron-positron pair-correlation
function (PCF) $g(r)$ and the annihilating-pair momentum density (APMD)
$\rho(\bar{p})$ constructed from the DFT orbitals are therefore poor.  The
contact PCF $g(0)$ is particularly important because it determines the
annihilation rate $\lambda = 3 g(0)/(4c^3r_s^3)$ \cite{krause_rehberg} for a
positron immersed in a paramagnetic HEG, where $r_s$ is the electron density
parameter and $c$ is the speed of light in vacuo \cite{rs_and_au}.  If the
electron and positron motions were uncorrelated $g(0)$ would be unity, but the
strong correlation leads to much larger values, particularly at low densities,
where an electron-positron bound state (positronium or Ps) or even an
electron-electron-positron bound state (Ps$^-$) may be formed.

We have used the variational and diffusion quantum Monte Carlo (VMC and DMC)
methods \cite{Ceperley_1980,foulkes_2001} as implemented in the
\textsc{casino} code \cite{casino_ref} to study a single positron in a HEG\@.
Fermionic antisymmetry is imposed via the fixed-node approximation, in which
the nodal surface is constrained to equal that of a trial wave function.  We
used Slater-Jastrow (SJ) and Slater-Jastrow-backflow (SJB) trial wave
functions \cite{Kwon_1998,Lopez_Rios_2006}.  The latter go beyond the
single-particle SJ nodal surface by replacing the particle coordinates in the
Slater determinants by ``quasiparticle coordinates.''  SJB wave functions give
the highest accuracy obtained to date for the HEG
\cite{Kwon_1998,Lopez_Rios_2006}.  We also tested two types of orbitals: (i)
plane-wave orbitals for each particle and (ii) orbitals which describe the
pairing between the electrons and positron.  The pairing orbitals were
obtained from mean-field calculations performed in the reference frame of the
positron, so the orbitals are functions of the separation of an electron
and the positron \cite{Drummond_2010}.  Within this impurity-frame DFT
(IF-DFT) method, the pairing orbitals describe the electron-positron
correlation quite well on their own \cite{Drummond_2010} and give a different
nodal surface from the plane-wave orbitals.  (NB, our QMC calculations
were performed in the laboratory frame.)  The four wave-function forms
used are:
\begin{eqnarray}
\Psi_{\rm PW}^{\rm SJ} & = & e^{J({\bf R})} \left[e^{i{\bf k}_i\cdot{\bf
r}_{\uparrow}}\right] \left[e^{i{\bf k}_j\cdot{\bf r}_{\downarrow}}\right]
\nonumber \\ \Psi_{\rm PW}^{\rm SJB} & = & e^{J({\bf R})} \left[e^{i{\bf
k}_i\cdot\left({\bf r}_{\uparrow}+\mbox{\boldmath $\xi$}({\bf
R})\right)}\right] \left[e^{i{\bf k}_j\cdot\left({\bf
r}_{\downarrow}+\mbox{\boldmath $\xi$}({\bf R})\right)}\right] \nonumber \\
\Psi_{\rm pair}^{\rm SJ} & = & e^{J({\bf R})} \left[\phi_i({\bf
r}_{\uparrow}-{\bf r}_{\rm p})\right] \left[\phi_j({\bf r}_{\downarrow}-{\bf
r}_{\rm p})\right] \nonumber \\ \Psi_{\rm pair}^{\rm SJB} & = & e^{J({\bf R})}
\left[\phi_i({\bf r}_{\uparrow}-{\bf r}_{\rm p}+\mbox{\boldmath $\xi$}({\bf
R}))\right] \left[\phi_j({\bf r}_{\downarrow}-{\bf r}_{\rm p}+\mbox{\boldmath
$\xi$}({\bf R}))\right], \nonumber \\ \label{eqn:trial_wave_functions}
\end{eqnarray}
where ${\bf R}$ denotes the positions of all the particles, ${\bf
r}_{\uparrow}$ and ${\bf r}_{\downarrow}$ denote the positions of up- and
down-spin electrons, respectively, ${\bf r}_{\rm p}$ is the positron position,
and $[\cdots]$ denotes a Slater determinant.  The Jastrow exponent $J({\bf
R})$ \cite{ndd_jastrow} and the backflow displacement $\mbox{\boldmath
$\xi$}({\bf R})$ \cite{Lopez_Rios_2006} contain parameters that were optimized
separately for each wave function and system.  The Jastrow exponents were
first optimized using the efficient VMC variance-minimization scheme of Ref.\
\onlinecite{Drummond_2005}, and then all the parameters (including the
backflow parameters) were optimized together using the VMC energy-minimization
scheme of Ref.\ \onlinecite{umrigar_emin}.  The pairing orbitals $\{\phi_i\}$
were represented using B-spline functions on a real-space grid
\cite{alfe_2004_blips}. The electron-positron cusp condition was enforced on
the pairing orbitals for wave function $\Psi_{\rm pair}^{\rm SJ}$
\cite{cusp_corr,binnie_lih}; for the other three wave functions, the cusp
conditions were imposed via the Jastrow factor.  In all our calculations the
simulation-cell Bloch vector \cite{rajagopal_kpoints} was chosen to be ${\bf
k}_s={\bf 0}$.

Tests at high ($r_s=1$) and low ($r_s=8$) electron densities show that the
qualitative features of the variations in $\Delta \Omega$, $g(r)$, and
$\rho(\bar{p})$ with $r_s$ are the same for each of the four wave functions of
Eq.\ (\ref{eqn:trial_wave_functions}).  However, as shown in the auxiliary
material \cite{EPAPS}, we obtained lower VMC and DMC energies with the SJB
wave functions ($\Psi_{\rm PW}^{\rm SJB}$ and $\Psi_{\rm pair}^{\rm SJB}$)
than the SJ ones ($\Psi_{\rm PW}^{\rm SJ}$ and $\Psi_{\rm pair}^{\rm SJ}$),
and therefore we used SJB wave functions to obtain all our main results.  The
pairing orbitals give lower SJB-VMC energies than the plane-wave orbitals, but
the SJB-DMC energies with the plane-wave and pairing orbitals are almost
identical.  The lack of sensitivity to the orbitals used, and hence the nodal
surface, suggests that the DMC energies are highly accurate.  The energies
reported in this paper are from DMC calculations using wave function
$\Psi_{\rm PW}^{\rm SJB}$.  Such calculations are considerably less expensive
than calculations using $\Psi_{\rm pair}^{\rm SJB}$ due to (i) the lower
energy variance achieved with $\Psi_{\rm PW}^{\rm SJB}$ \cite{EPAPS} and (ii)
the fact that plane-wave orbitals are cheaper to evaluate.  The DMC energies
were extrapolated to zero time step.  Our production DMC calculations were
performed in cells containing $N=54$ electrons.  Tests of convergence with
respect to system size up to $N=114$ electrons are described in the auxiliary
material \cite{EPAPS}. The cell volume was chosen to be $(N-1)\times(4/3)\pi
r_s^3$, so that the electron density far from the positron was correct.
IF-DFT calculations \cite{Drummond_2010} suggest that finite-size effects due
to the interaction of images of the positron are negligible for $N \geq 54$
electrons.

\begin{figure}
\begin{center}
\includegraphics[clip,scale=0.33]{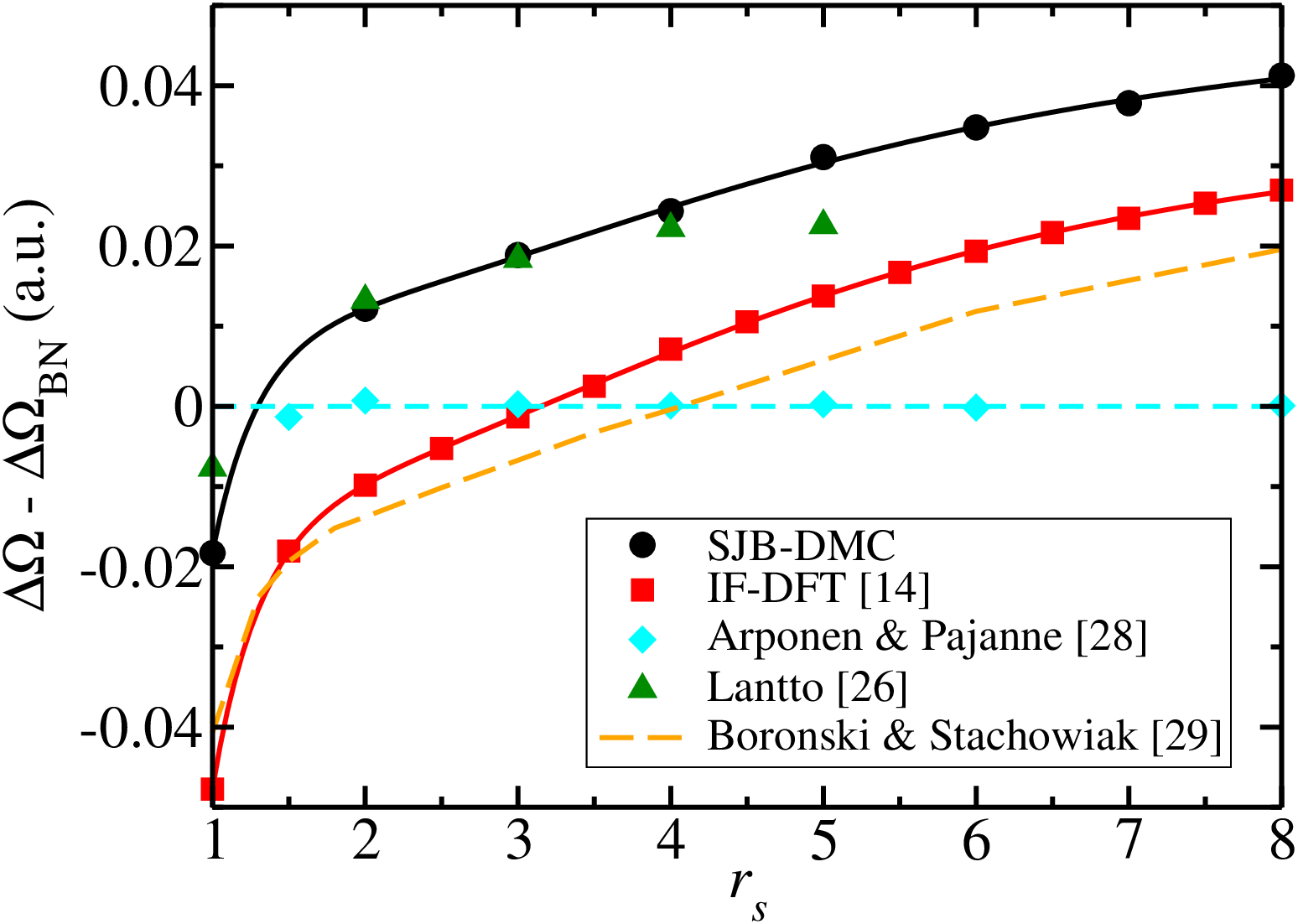}
\caption{(Color online) Relaxation energy against density parameter from our
  SJB-DMC calculations and other studies
  \cite{lantto_1987,arponen_1979,boronski_1998,Drummond_2010}, relative to the
  Boro\'nski-Nieminen expression, $\Delta \Omega_{\rm BN}$
  \cite{boronski_1986} (horizontal dashed line).
\label{fig:final_relax_rel}}
\end{center}
\end{figure}

Our DMC relaxation energies are plotted in Fig.\ \ref{fig:final_relax_rel} and
are well-fitted by the form
\begin{equation}
  \Delta \Omega(r_s) = \frac{A_{-1} r_s^{-1} +
    A_0+A_1r_s-0.262005B_2r_s^2}{1+B_1r_s+B_2r_s^2},
\label{eq:relaxation_energy}
\end{equation}
where $A_{-1}=-0.260361$, $A_0=-0.261762$, $A_1=0.00375534$, $B_1=0.113718$,
and $B_2=0.0270912$. Equation (\ref{eq:relaxation_energy}) tends to the
correct low-density limit of the energy of the Ps$^-$ ion \cite{frolov_2005}.
Equation (\ref{eq:relaxation_energy}) does not yield the exact high-density
behavior of the random phase approximation (RPA), although this is only
relevant for $r_s<0.1$ \cite{arponen_1978}.  VMC energies for a positron in a
HEG have been reported previously \cite{ortiz_1992}, but we have used superior
trial wave functions and have obtained very different results.  At high
densities our relaxation energies are similar to those of Lantto
\cite{lantto_1987}, but at lower densities we obtain smaller values.  The
SJB-DMC and IF-DFT results \cite{Drummond_2010} and the data of Boro\'nski and
Stachowiak \cite{stachowiak_1990} show similar behavior with $r_s$, while the
Boro\'nski-Nieminen fit \cite{boronski_1986} to the data of Ref.\
\onlinecite{arponen_1979} is markedly different.  Boro\'nski and Nieminen's
\cite{boronski_1986} expression for $\Delta \Omega(r_s)$ is widely used in
two-component DFT calculations, but our study suggests it is not very accurate
and should be replaced by Eq.\ (\ref{eq:relaxation_energy}).

\begin{figure}
\begin{center}
\includegraphics[clip,scale=0.33]{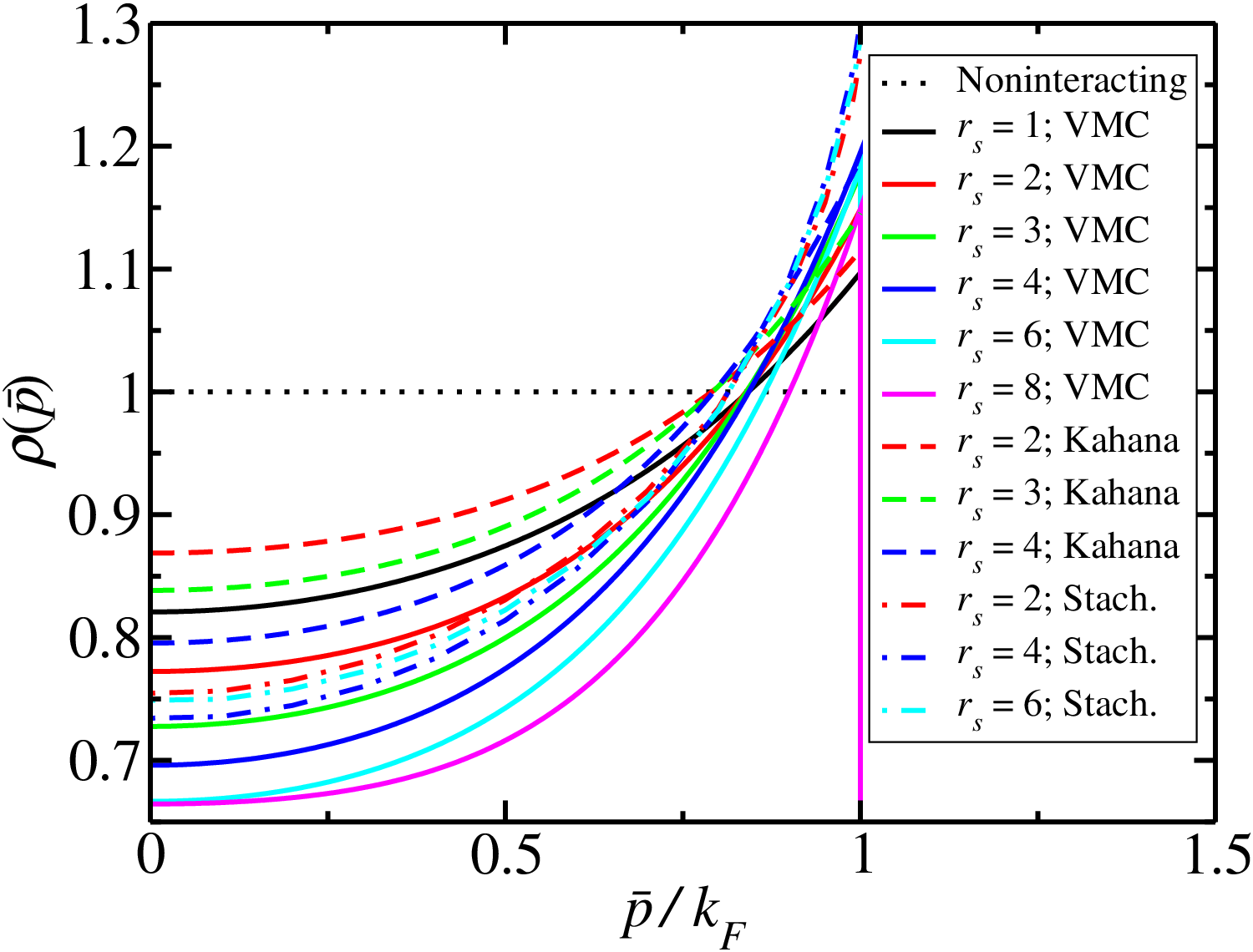}
\includegraphics[clip,scale=0.33]{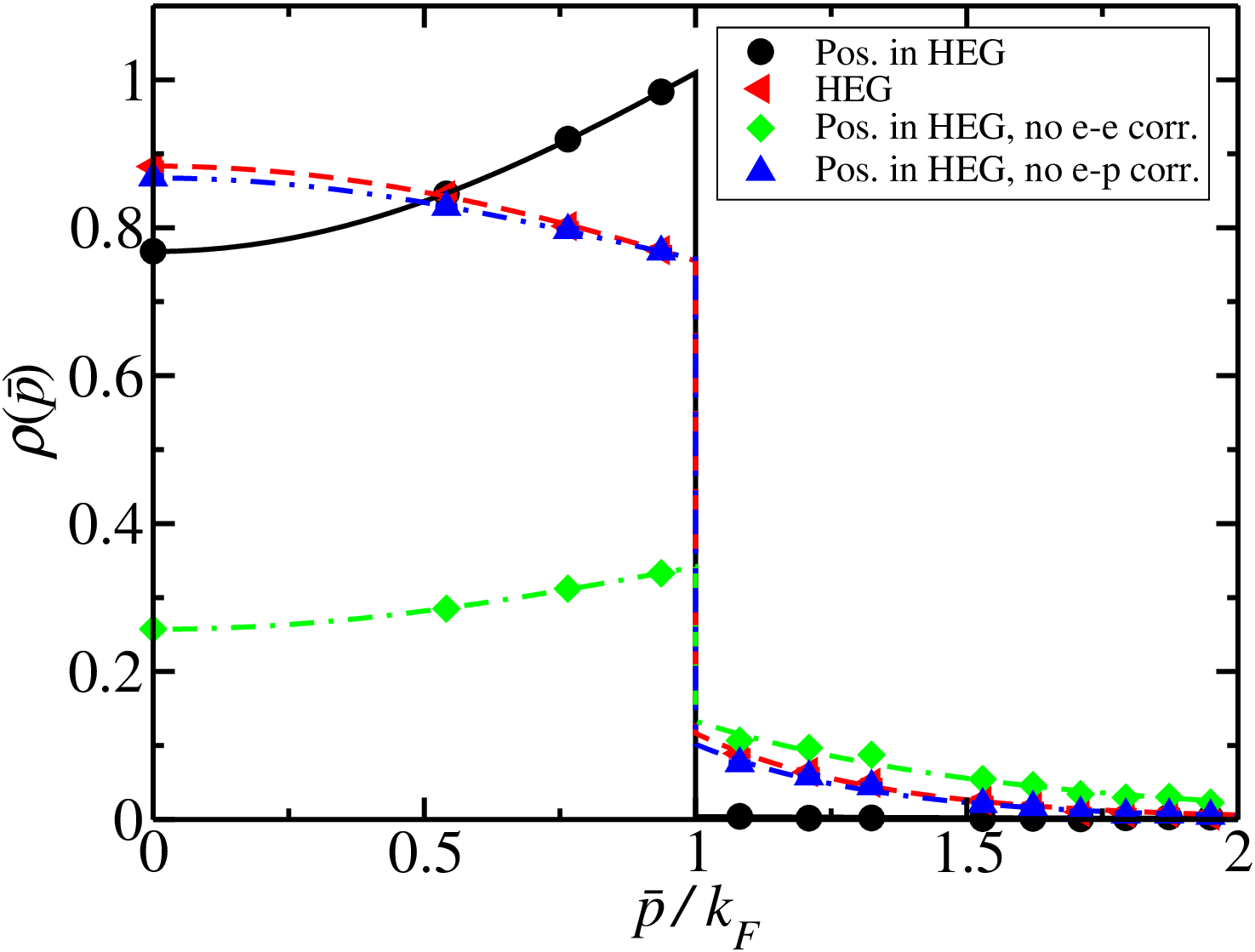}
\caption{(Color online) Top: APMDs [$\rho(\bar{p})$] for different densities.
The solid lines show our VMC data obtained with wave function $\Psi_{\rm
pair}^{\rm SJB}$ and $N=114$ electrons, while the dashed and dashed-dotted
lines show the data of Kahana \cite{kahana_1963} and Stachowiak
\cite{stachowiak_1990}, respectively.  Bottom: APMDs for the positron-in-HEG
and the HEG at $r_s=8$ and $N=54$ electrons, calculated using $\Psi_{\rm
PW}^{\rm SJB}$.
\label{fig:final_md}}
\end{center}
\end{figure}

We calculated the APMD within VMC using optimized SJB trial wave functions
with pairing orbitals ($\Psi_{\rm pair}^{\rm SJB}$), because these give lower
VMC energies than plane-wave orbitals ($\Psi_{\rm PW}^{\rm SJB}$).  These
calculations were performed by constraining an electron and the positron to
lie on top of one another throughout the simulation \cite{EPAPS}.  APMDs at
different densities are plotted in Fig.\ \ref{fig:final_md}, with the
normalization chosen such that $\int_0^\infty 4 \pi \bar{p}^2 \rho(\bar{p})\,
d\bar{p} = (4/3) \pi k_F^3$.  Our results clearly show the enhancement of the
APMD below the Fermi momentum predicted by Kahana \cite{kahana_1963}, but our
data differ quantitatively from previous results
\cite{kahana_1963,stachowiak_1990,Drummond_2010}.  Our VMC data have almost no
weight above the Fermi momentum over the entire density range studied, even
though the weight in the MD above $k_F$ in the HEG is substantial at low
densities.  For example, we find that the APMD immediately above $k_F$ is
roughly 10\% of the value for the HEG at $r_s=1$ and 3\% at $r_s=8$.

Suppression of the weight in the APMD above $k_F$ was demonstrated
theoretically by Carbotte and Kahana \cite{Carbotte_1965}, but our study gives
a more detailed and accurate picture.  We investigated the weight above $k_F$
using VMC with the wave function $\Psi_{\rm PW}^{\rm SJB}$ by selectively
eliminating interparticle correlations.  Neglecting electron-electron and
electron-positron correlations gives the familiar ``top hat'' MD of the
noninteracting system.  Calculations with the electron-positron terms removed
give an APMD indistinguishable from the MD of the HEG, with a tail above
$k_F$.  Calculations including electron-positron correlation but neglecting
electron-electron correlation show Kahana enhancement below $k_F$ and a tail
above $k_F$.  When, however, both electron-electron and electron-positron
correlations are included, the tail above $k_F$ is largely suppressed, as
shown in the lower panel of Fig.\ \ref{fig:final_md}.

The suppression of the tail in the APMD can be explained by examining the
behavior of the two-body terms in the Jastrow exponent.  (For simplicity, we
consider the $\Psi_{\rm PW}^{\rm SJ}$ wave function in the following
discussion.)  The Jastrow exponent $J({\bf R})$ is the sum of
electron-electron [$u_{\uparrow\uparrow}(r)$ and $u_{\uparrow\downarrow}(r)$,
where the arrows indicate spins] and electron-positron [$u_{\rm ep}(r)$]
terms.  If one assumes that
\begin{eqnarray}
  u_{\uparrow\downarrow}(r) = u_{\uparrow\uparrow}(r) = -u_{\rm ep}(r),
\label{equality_of_jastrows}
\end{eqnarray}
then the APMD has exactly zero weight above $k_F$, as shown in the auxiliary
material \cite{EPAPS}. The RPA (linear response theory) shows that Eq.\
(\ref{equality_of_jastrows}) holds at large $r$ and the Kato cusp conditions
force the gradients of $u_{\uparrow\downarrow}(r)$ and $u_{\rm ep}(r)$ to
satisfy Eq.\ (\ref{equality_of_jastrows}) at $r=0$.  The cusp conditions for
parallel and antiparallel spin electrons are different and therefore
$u_{\uparrow\downarrow}(r)$ and $u_{\uparrow\uparrow}(r)$ must differ at small
$r$, but antisymmetry ensures that the probability of parallel-spin electrons
being closer than $r_s$ is small.  As shown in the auxiliary material, plots
of the terms in the Jastrow exponent demonstrate the approximate validity of
Eq.\ (\ref{equality_of_jastrows}).

\begin{figure}[ht!]
\includegraphics[clip,scale=0.33]{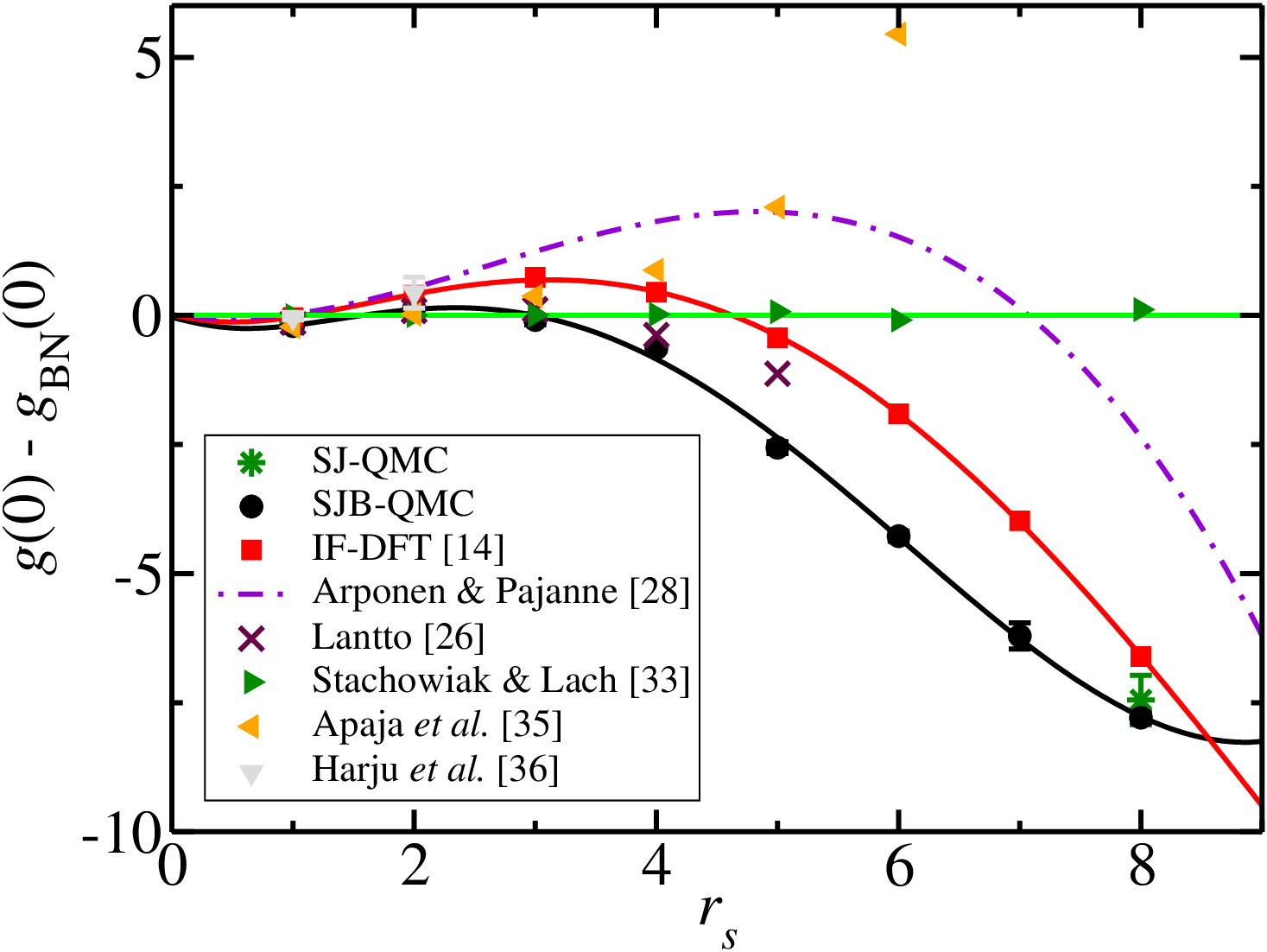}
\caption{(Color online) Deviation of the contact PCF $g(0)$ from the form
$g_{\rm BN}(0)$ of Boro\'nski and Nieminen \cite{boronski_1998} together with
other results in the literature
\cite{Drummond_2010,arponen_1979,lantto_1987,stachowiak_1993,apaja_2003,Harju_1996}.
\label{fig:g0_relative}}
\end{figure}

We calculated the PCFs within VMC and DMC using $\Psi_{\rm PW}^{\rm SJB}$ wave
functions, because these give the same results as pairing orbitals but the
calculations are much cheaper \cite{EPAPS}.  The final results were evaluated
by extrapolated estimation (twice the DMC PCF minus the VMC PCF)
\cite{Ceperley_1986}, in order to eliminate the leading-order errors.  In
Fig.\ \ref{fig:g0_relative}, the electron-positron contact PCF $g(0)$ is
plotted relative to the Boro\'nski-Nieminen form \cite{boronski_1998}, which
is a fit to the data of Stachowiak and Lach \cite{stachowiak_1993}. Our
contact PCF data are well-represented by
\begin{eqnarray}
  g(0) & = & 1+1.23r_s+a_{3/2}r_s^{3/2}+a_2r_s^2 + a_{7/3}r_s^{7/3} \nonumber
  \\ & & {} +a_{8/3}r_s^{8/3}+0.173694r_s^3,
\label{eqn:pcf_fit}
\end{eqnarray}
where $a_{3/2}=-3.38208$, $a_2=8.6957$, $a_{7/3}=-7.37037$, and
$a_{8/3}=1.75648$.  Equation (\ref{eqn:pcf_fit}) satisfies the high-density
(RPA) \cite{arponen_1978} and low-density (Ps$^-$) limiting behaviors
\cite{frolov_2005}.  Our full data for $g(r)$ are given in the auxiliary
material \cite{EPAPS}.  The IF-DFT data follow the extrapolated SJB data quite
well, while the other many-body calculations give somewhat larger values of
$g(r)$ at low densities.  In the density range $r_s=5$--8 a.u., our values of
$g(0)$ are approximately 9\% smaller than those given by the
Boro\'nski-Nieminen expression \cite{boronski_1986}.  The local increase of
the electron density around the positron caused by their mutual attraction is
modeled in two-component DFT using an ``enhancement factor'' based on data for
$g(0)$.  Using our smaller values of $g(0)$ would reduce the enhancement
factor and hence the overestimation of annihilation rates obtained with the
positronic LDA \cite{Mitroy_2002}.

In conclusion, our results are the most accurate obtained so far for a
positron in a HEG\@.  Our data for $\Delta \Omega$ are sufficient to define
the energy functional for a two-component positronic DFT within the
LDA\@. They would also be useful in developing semilocal
\cite{Barbiellini_1996} or other functionals.  Our PCF data give a smaller
enhancement factor than the standard Boro\'nski-Nieminen expression
\cite{boronski_1986}.  Our APMDs have very little weight above $k_F$ because
of the cancellation of electron-electron and electron-positron correlation
effects.  We have derived an exact result relating Eq.\
(\ref{equality_of_jastrows}) to the complete absence of weight in the APMD
$\rho(\bar{p})$ for $\bar{p}>k_F$, which is useful in understanding this
effect.

\begin{acknowledgments}
  We acknowledge financial support from the UK Engineering and Physical
  Sciences Research Council (EPSRC)\@.  Computer resources were provided by
  the Cambridge High-Performance Computing Facility and the Lancaster High-End
  Computing cluster.
\end{acknowledgments}


\begin{thebibliography}{99}

\bibitem{krause_rehberg} R.\ Krause-Rehberg and H.S.\ Leipner,
  \textit{Positron Annihilation in Semiconductors} (Springer-Verlag, Berlin,
  1999).

\bibitem{Gribakin_2010} G.F.\ Gribakin, J.A.\ Young, and C.M.\ Surko, Rev.\
  Mod.\ Phys.\ \textbf{82}, 2557 (2010).

\bibitem{Pethrick_1997} R.A.\ Pethrick, Prog.\ Polym.\ Sci.\ \textbf{22}, 1
  (1997).

\bibitem{Gidley_2006} D.W.\ Gidley, H.-G.\ Peng, and R.S.\ Vallery, Annu.\
  Rev.\ Mater.\ Res.\ \textbf{36}, 49 (2006).

\bibitem{Major_2004} Zs.\ Major, S.B.\ Dugdale, R.J.\ Watts, G.\ Santi, M.A.\
  Alam, S.M.\ Hayden, J.A.\ Duffy, J.W.\ Taylor, T.\ Jarlborg, E.\ Bruno, D.\
  Benea, and H.\ Ebert, Phys.\ Rev.\ Lett.\ \textbf{92}, 107003 (2004).

\bibitem{boronski_1986} E.\ Boro\'nski and R.M.\ Nieminen, Phys.\ Rev.\ B
  \textbf{34}, 3820 (1986).

\bibitem{puska_1994} M.J.\ Puska and R.M.\ Nieminen, Rev.\ Mod.\ Phys.\
  \textbf{66}, 841 (1994).

\bibitem{rs_and_au} We use Hartree atomic units ($\hbar=|e|=m_{\rm
    e}=4\pi\epsilon_0=1$) throughout.

\bibitem{Ceperley_1980} D.M.\ Ceperley and B.J.\ Alder, Phys.\ Rev.\ Lett.\
  \textbf{45}, 566 (1980).

\bibitem{foulkes_2001} W.M.C.\ Foulkes, L.\ Mitas, R.J.\ Needs, and G.\
  Rajagopal, Rev.\ Mod.\ Phys.\ \textbf{73}, 33 (2001).

\bibitem{casino_ref} R.J.\ Needs, M.D.\ Towler, N.D.\ Drummond, and P.\
  L\'opez R\'ios, J.\ Phys.:\ Condens.\ Matter \textbf{22}, 023201 (2010).

\bibitem{Kwon_1998} Y.\ Kwon, D.M.\ Ceperley, and R.M.\ Martin, Phys.\ Rev.\ B
  \textbf{58}, 6800 (1998).

\bibitem{Lopez_Rios_2006} P.\ L\'opez R\'{\i}os, A.\ Ma, N.D.\ Drummond, M.D.\
  Towler, and R.J.\ Needs, Phys.\ Rev.\ E \textbf{74}, 066701 (2006).

\bibitem{Drummond_2010} N.D.\ Drummond, P.\ L\'opez\ R\'{\i}os, C.J.\ Pickard,
  and R.J.\ Needs, Phys.\ Rev.\ B \textbf{82}, 035107 (2010).

\bibitem{ndd_jastrow} N.D.\ Drummond, M.D.\ Towler, and R.J.\ Needs, Phys.\
  Rev.\ B \textbf{70}, 235119 (2004).

\bibitem{Drummond_2005} N.D.\ Drummond and R.J.\ Needs Phys.\ Rev.\ B
  \textbf{72}, 085124 (2005).

\bibitem{umrigar_emin} C.J.\ Umrigar, J.\ Toulouse, C.\ Filippi, S.\ Sorella,
  and R.G.\ Hennig, Phys.\ Rev.\ Lett.\ \textbf{98}, 110201 (2007).

\bibitem{alfe_2004_blips} D.\ Alf\`{e} and M.J.\ Gillan, Phys.\ Rev.\ B
\textbf{70}, 161101 (2004).

\bibitem{cusp_corr} A.\ Ma, M.D.\ Towler, N.D.\ Drummond, and R.J.\ Needs, J.\
Chem.\ Phys.\ \textbf{122}, 224322 (2005).

\bibitem{binnie_lih} S.J.\ Binnie, S.J.\ Nolan, N.D.\ Drummond, D.\ Alf\`{e},
  N.L.\ Allan, F.R.\ Manby, and M.J.\ Gillan, Phys.\ Rev.\ B \textbf{82},
  165431 (2010).

\bibitem{rajagopal_kpoints} G.\ Rajagopal, R.J.\ Needs, A.\ James, S.D.\
  Kenny, and W.M.C.\ Foulkes, Phys.\ Rev.\ B \textbf{51}, 10591 (1995).

\bibitem{EPAPS} See EPAPS Document No.\ xxxxxxxxxx. For more information on
  EPAPS, see http://www.aip.org/pubservs/epaps.html.

\bibitem{frolov_2005} A.M.\ Frolov, Phys.\ Lett.\ A \textbf{342}, 430 (2005).

\bibitem{arponen_1978} J.\ Arponen, J.\ Phys.\ C \textbf{11}, L739 (1978).

\bibitem{ortiz_1992} G.\ Ortiz, PhD thesis, Swiss Federal Institute of
  Technology, Lausanne (1992).  The relevant data from this thesis are also
  reported in Refs.\ \onlinecite{apaja_2003} and \onlinecite{boronski_1998}.

\bibitem{lantto_1987} L.J.\ Lantto, Phys.\ Rev.\ B \textbf{36}, 5160 (1987).

\bibitem{stachowiak_1990} H.\ Stachowiak, Phys.\ Rev.\ B \textbf{41}, 12522
  (1990).

\bibitem{arponen_1979} J.\ Arponen and E.\ Pajanne, Ann.\ Phys.\ \textbf{121},
  343 (1979).

%Gives Ortiz's correlation data in Figs. 2 and 3 and Table 3.
%Positron-electron correlation energy in an electron gas according to
%the perturbed-hypernetted-chain approximation
\bibitem{boronski_1998} E.\ Boro\'nski and H.\ Stachowiak, Phys.\ Rev.\ B
  \textbf{57}, 6215 (1998).

\bibitem{kahana_1963} S.\ Kahana, Phys.\ Rev.\ \textbf{129}, 1622 (1963).

\bibitem{Carbotte_1965} J.P.\ Carbotte and S.\ Kahana Phys.\ Rev.\
  \textbf{139}, A213 (1965).

\bibitem{Ceperley_1986} D.M.\ Ceperley and M.H.\ Kalos, in \textit{Monte Carlo
  Methods in Statistical Physics} 2nd edn, edited by K.\ Binder
  (Springer-Verlag, Heidelberg, 1979), p.\ 145.

\bibitem{stachowiak_1993} H.\ Stachowiak and J.\ Lach, Phys.\ Rev.\ B
\textbf{48}, 9828 (1993).

\bibitem{Mitroy_2002} J.\ Mitroy and B.\ Barbiellini, Phys.\ Rev.\ B
\textbf{65}, 235103 (2002).

%Ortiz's contact PCF included in Table I
\bibitem{apaja_2003} V.\ Apaja, S.\ Denk, and E.\ Krotscheck, Phys.\ Rev.\ B
\textbf{68}, 195118 (2003).

\bibitem{Harju_1996} A.\ Harju, B.\ Barbiellini, S.\ Siljam\"aki, R.M.\
  Nieminen, and G.\ Ortiz, J.\ Radioanal.\ Nucl.\ Chem.\ \textbf{211}, 193
  (1996).

\bibitem{Barbiellini_1996} B.\ Barbiellini, M.J.\ Puska, T.\ Korhonen, A.\
  Harju, T.\ Torsti, and R.M.\ Nieminen, Phys.\ Rev.\ B \textbf{53}, 16201
  (1996).

\end{thebibliography}
\end{document}